\documentclass[12pt,a4paper,notitlepage,aps]{revtex4}
\usepackage{latexsym}
\newcommand\beq{\begin{equation}}
\newcommand\eeq{\end{equation}}
\newcommand\bea{\begin{eqnarray}}
\newcommand\eea{\end{eqnarray}}
\begin{document}
\title{ Irreducible SU(3) Schwinger Bosons} 

\author{Ramesh Anishetty\footnote{e-mail: ramesha@imsc.res.in}}
\affiliation{The Institute of Mathematical Sciences, CIT-Campus, Taramani,
Chennai, India}

\author{Manu Mathur\footnote{e-mail: manu@bose.res.in} and Indrakshi Raychowdhury\footnote{e-mail: indrakshi@bose.res.in}}
\affiliation{S. N. Bose National Centre 
for Basic Sciences, JD Block, Sector III, Salt Lake City,  Kolkata 
700098, India. }


\begin{abstract}


\noindent We develop simple computational techniques for constructing all possible SU(3) 
representations in terms of irreducible SU(3) Schwinger bosons. We show that these irreducible 
Schwinger oscillators make SU(3) representation theory as simple as SU(2).  
The new Schwinger oscillators satisfy certain Sp(2,R) constraints and solve the 
multiplicity problem as well. These SU(3) techniques can be generalized to SU(N).  
  
\end{abstract}
\maketitle

\section{Introduction}

It is well known that the simple features of SU(2) Lie algebra or angular 
momentum algebra and it's representations are lost when we study SU(3) or 
higher SU(N) groups. In the past, considerable work has been done in this 
 direction \cite{apw,sharat,mm1,mu1}. A detailed discussion with an 
extensive list of references in this context can be found in \cite{sharat}.
The purpose and motivation of this work is to address these issues and make 
the construction of all SU(3) irreducible representations (irreps.) as simple 
and accessable as SU(2). Moreover, the techniques can be  generalized to higher 
SU(N) groups in a straightforward manner. We use Schwinger boson 
representation of SU(2) \cite{schwinger} and SU(3) Lie algebra to illustrate our results. 
Infact,  Schwinger boson representation of SU(2) Lie algebra  has played 
many important and diverse roles  in different areas of physics due to its intrinsic 
simplicity. More explicitly, the simplicity is because the Schwinger  analysis 
of the SU(2) Lie algebra  is in terms of it's  smallest (spin half) constituents   
instead of (spin one) angular momentum operators themselves. These spin half operators are 
two quantum mechanical harmonic oscillators $(j= {1 \over 2}, m = \pm {1 \over 2})$ 
which are called Schwinger bosons.   
The angular momentum operators  are composites of Schwinger bosons and belong to 
(higher) spin one representation. 
This leads to enormous computational simplifications in the representation theory 
of SU(2) as well as it's various coupling coefficients \cite{schwinger}. Besides this 
group theory advantages, the Schwinger SU(2) construction has also 
been exploited in nuclear physics \cite{np}, strongly correlated systems \cite{scs}, 
supersymmetry and supergravity algebras \cite{susy}, lattice gauge theories \cite{lgt}, 
loop quantum gravity \cite{lqg} etc.

In the context of SU(2) group theory, the Schwinger bosons, each carrying basic (half) unit 
of spin angular momentum flux, provide an explicit and simple realization of the angular momentum 
algebra as well as all it's representations \cite{schwinger}. 
In particular, the Hilbert space  created by the two Schwinger oscillators  is 
isomorphic to the representation space of SU(2) group (see section II). Thus the 
Schwinger boson representation of SU(2) group is simple,  economical as well as 
complete. However,  all these  features are lost when we 
generalize the Schwinger boson construction to SU(N) with $N \ge 3$ (see section III). 
The origin of  these problems is the existence of certain SU(N) invariants 
which can be constructed for $N \ge 3$. Any two states   
which differ by the overall presence of such an invariants will transform in the 
same way under SU(N). This leads to the problem of multiplicity which in turn 
makes the representation theory of SU(N) ($N\ge 3$) much more involved compared 
to SU(2) (compare representations (\ref{su2_irrep}) for SU(2) with (\ref{bv}) for 
SU(3)). 
These issues have also been extensively addressed in the past  
\cite{apw,sharat,mm1,mu1}. In the context of SU(3) Schwinger boson analysis, a systematic 
group theoretic procedure based on noncompact group Sp(2,R) is given in \cite{mu1} to 
label the multiplicity of SU(3). In this work, exploiting this Sp(2,R) labeling in 
\cite{mu1}, we define irreducible SU(3) Schwinger bosons in terms of 
which construction of SU(3) representations is as simple as SU(2) 
(compare (\ref{su2_irrep}) for SU(2) with (\ref{modsu3irrep}), instead of 
(\ref{bv}),  for SU(3)).  Further like in SU(2) case, the 
representations in terms of irreducible Schwinger bosons are multiplicity free. 

The plan of the paper is as follows. In section II, we start with a brief 
introduction to SU(2)  Schwinger bosons. This section  makes the presentation 
 self contained and also allows us to compare our SU(3) results (section IV) 
with the corresponding SU(2) results explicitly. In section III we discuss 
SU(3) Lie algebra and it's representations in terms of Schwinger bosons. 
We also summarize the construction of Sp(2,R) group in \cite{mu1} which commutes 
with SU(3). In section IV, we solve certain Sp(2,R) constraints in terms of irreducible 
Schwinger bosons leading to all SU(3) representations with the simplicity of SU(2). 
The appendix A describes the construction of SU(3) projection operators to construct 
the SU(3) irreps. discussed in section III. In appendix B we show that the  
irreducible Schwinger bosons in $(1,0)$ and $(0,1)$ representations acting on 
$(n,m)$ SU(3) irrep. directly produce $(n+1,m)$ and $(n,m+1)$ irreps. respectively.

\section{SU(2) Schwinger Bosons}

The $SU(2)$ Lie 
algebra is given by a set of three angular momentum operators $\lbrace \vec{J} 
\rbrace \equiv \lbrace J_1,J_2,J_3 \rbrace$ or equivalently by $\lbrace J_{+},
J_{-},J_{3} \rbrace$, $(J_{\pm} \equiv J_{1} \pm i~ J_{2})$ satisfying
\beq
[J_{3},J_{\pm}] = \pm J_{\pm}, ~~~~~~ [J_{+},J_{-}] = 2 J_{3} ~.
\label{su2a} 
\eeq

\noindent The $SU(2)$ group has a Casimir operator given by $\vec{J} \cdot 
\vec{J}$, and the different irreducible representations are characterized 
by its eigenvalues $j(j+1)$, where $j$ is an integer or half-odd-integer. 
A given basis vector in representation $j$ is labeled by the eigenvalue $m$ 
of $J_3$ as $\vert j,m\rangle$.

\noindent 
We now define a doublet of quantum mechanical oscillators or equivalently Schwinger bosons,  
$\vec{a} \equiv (a_1, a_2)$ and $ \vec{a}^{\dagger} \equiv (a^{\dagger 1}, 
a^{\dagger 2})$ respectively \cite{schwinger}. 
They satisfy the simpler bosonic commutation relation
$[a_\alpha,a^{\dagger \beta}] = \delta_{\alpha}^{\beta}$ with $\alpha, \beta =1,2$. The angular momentum operators in (\ref{su2a})  
are constructed out of Schwinger bosons as: 
\beq
J^a \equiv  a^{\dagger}\frac{\sigma^{\mathrm a}}{2} a,
\label{sch} 
\eeq
where $\sigma^{\mathrm a}$ denote the Pauli matrices.
It is easy to check that 
the operators in (\ref{sch}) satisfy the $SU(2)$ Lie algebra with the SU(2) 
Casimir:  
\bea 
\vec{J} \cdot \vec{J} \equiv {\vec{a}^{\dagger} \cdot \vec{a} \over 2}  
\left({\vec{a}^{\dagger} \cdot \vec{a} \over 2} + 1\right).\eea 
Thus the representations of $SU(2)$ 
can be characterized by the eigenvalues of the total occupation number operator with the angular 
momentum satisfying,
\bea j={\left(n_1+ n_2\right) \over 2}  \equiv {n \over 2}   \eea
where $n_1$ and $n_2$ are the 
eigenvalues of $a^{\dagger 1} a_1$ and $a^{\dagger 2} a_2$ respectively. \\
A general irreducible representation of $SU(2)$ with $ n = n_1+n_2=2j$ can be written as,
\beq
\label{su2_irrep}
|\psi\rangle^{\alpha_1\alpha_2...\alpha_{n}} 
=a^{\dagger \alpha_1} a^{\dagger \alpha_2}... a^{\dagger \alpha_{n}}|0\rangle
\equiv O^{\alpha_1\alpha_2 \ldots \alpha_n} |0\rangle.  
\eeq
Above  we have defined $O^{\alpha_1\alpha_2 \ldots \alpha_n} = 
a^{\dagger \alpha_1} a^{\dagger \alpha_2}... a^{\dagger \alpha_{n}}$ for later convenience.
The state $|0\rangle$ denotes the vaccum state of both the Schwinger bosons, i.e., 
$a_\alpha|0\rangle = 0, ~\alpha=1,2$.  
The states in (\ref{su2_irrep}) are completely symmetric in all the $n = 2j$ number of $\alpha$ indices.
Graphically, the representation (\ref{su2_irrep}) is given by Young 
tableau in Fig. 1. 
\begin{figure}[h] 
\setlength{\unitlength}{3947sp}%
\begingroup\makeatletter\ifx\SetFigFont\undefined%
\gdef\SetFigFont#1#2#3#4#5{%
  \reset@font\fontsize{#1}{#2pt}%
  \fontfamily{#3}\fontseries{#4}\fontshape{#5}%
  \selectfont}%
\fi\endgroup%
\begin{center} 
\begin{picture}(1824,627)(1189,-976)
\thinlines
{\put(1201,-661){\framebox(300,300){$\alpha_1$}}
}%
{\put(1501,-661){\framebox(300,300){$\alpha_2$}}
}%
{\put(2701,-661){\framebox(300,300){$\alpha_n$}}
}%
{\put(1801,-661){\framebox(900,300){$\ldots$}}
}%
\put(1726,-961){\makebox(0,0)[lb]{\smash{{\SetFigFont{12}{14.4}{\rmdefault}{\mddefault}{\updefault}{$n=2j$ boxes}%
}}}}
\end{picture}%
\caption[su2irrep]{The SU(2) representation in terms of Young Tableau.  Each 
Schwinger boson $a^{\dagger \alpha}$ in (\ref{su2_irrep}) creates a Young tableau box.} 
\end{center} 
\end{figure} 

As mentioned in the introduction, the aim of 
the present work is to define irreducible SU(3) Schwinger bosons in terms of which all 
SU(3) representations retain the simplicity of (\ref{su2_irrep}).  However, before going 
into technical details, we briefly summarize the essential preliminary ideas of SU(3) Lie 
algebra and it's representations in terms of SU(3) Schwinger bosons. 

\section{SU(3) Schwinger Bosons}

The rank of the $SU(3)$ group is two. Therefore, to cover all SU(3) irreducible representations  
we  need two independent harmonic oscillator triplets belonging to the two 
fundamental representations $3$ and $3^{*}$. 
Let's denote them by $\{a^{\dagger}_\alpha\}$ and $\{b^{\dagger\alpha}\}$ respectively with 
$\alpha = 1,2,3$. Now the generator of $SU(3)$ group are written as \cite{georgi}:
\bea
Q^{\mathrm a} = a^\dagger {\lambda^{\mathrm a} \over 2}  a -
b^\dagger {\tilde{\lambda}^{\mathrm a} \over 2} b
\label{su3sb} 
\eea
In (\ref{su3sb}), ${\mathrm a} =1,2,...,8$, $\lambda^{\mathrm a}$ are the Gell Mann matrices for the 
triplet ($3$) representation,  $- \tilde{\lambda}^{\mathrm a}$ are the corresponding matrices 
for the anti-triplet $3^*$ representation where $\tilde{\lambda}$ denotes the transpose 
of $\lambda$. Throughout the paper, the upper and lower 
indices are in the conjugate $3$ (i.e,  transforming as $a^{\dagger\alpha}$) 
and $3^{*}$ (i.e, transforming as  $b^{\dagger}_{\alpha}$)  
representations  respectively. 
The operators $Q^a$ satisfy the $SU(3)$ algebra amongst themselves, i.e, $\quad[Q^a,Q^b]=if^{abc}Q^c$ where 
$f^{abc}$ are the SU(3) structure constants \cite{georgi}. 
The defining relations (\ref{su3sb}) also imply: 
\bea
~~~\big[Q^{\mathrm a}, (a^{\dagger})^{\alpha} \big] = 
{1 \over 2} 
(a^{\dagger})^\beta 
\left({\lambda^{\mathrm a}}\right)_{\beta}^{~\alpha},~~~~~~~ 
\big[Q^{\mathrm a}, {b^{\dagger}}_{\alpha}\big]  =  - {1 \over 2} 
(\lambda^{\mathrm a})_{\alpha}^{~\beta}  {b^{\dagger}}_{\beta}. 
\label{cas} 
\eea
Thus the operators $(a^{\dagger})_{\alpha}$ and $(b^{\dagger})^{\alpha}$ transform according to 
$3$ and $3^*$ representations. The corresponding Young tableau are given by one box and two 
vertical boxes respectively.                   
As  $Q^{\mathrm a}, ({\mathrm a}=1,2,..,8)$ in (\ref{su3sb}) involve both creation 
and annihilation operators  
of $a$ and $b$ types, it is clear that:
\bea 
~~~~~\big[Q^{\mathrm a},{N}_a] =  0, ~~~~~~~~~~\big[Q^{\mathrm a},{N}_b\big] = 0. 
\label{comr}
\eea
The  SU(3) Casimirs are given by the total 
occupation numbers of $a$  and $b$  type oscillators: 
\bea
{N}_a = a^\dagger \cdot a,   \qquad \qquad {N}_b =  b^\dagger \cdot b.
\label{su3cas} 
\eea
We represent their eigenvalues by $n$ and $m$ respectively and the SU(3) vacuum $(n=0,~m=0)$ by the state $|0\rangle$.

\subsection{\bf Symmetric vs. Mixed representations} 

A general SU(3) irreducible representation is characterized by $(n,m)$. Note that n and m are 
the number of single and double boxes in a Young tableau diagram. At this stage, it is convenient 
to define the most basic SU(3) tensor operator: 
\bea 
O^{\alpha_1\alpha_2...\alpha_n}_{\beta_1\beta_2...\beta_m} \equiv
(a^{\dagger})^{\alpha_1} (a^{\dagger})^{\alpha_2} ... (a^{\dagger})^{\alpha_n}  
(b^{\dagger})_{\beta_1} (b^{\dagger})_{\beta_2} ... (b^{\dagger})_{\beta_m}, 
\label{defo} 
\eea 
with $O^{\alpha_1\alpha_2...\alpha_n} \equiv (a^{\dagger})^{\alpha_1} 
(a^{\dagger})^{\alpha_2} ... (a^{\dagger})^{\alpha_n}$ and 
$O_{\beta_1\beta_2...\beta_m} \equiv (b^{\dagger})_{\beta_1} 
(b^{\dagger})_{\beta_2} ... (b^{\dagger})_{\beta_m}$. 
A general irreducible $(n,m)$ representation of 
$SU(3)$, denoted by $|\psi\rangle ^{\alpha_1\alpha_2\ldots\alpha_n}_{\beta_1\beta_2\ldots\beta_m}$, 
satisfies the following three conditions \cite{Coleman}: \\ 
\phantom{x} {\bf C1:} symmetry in all upper $(\alpha)$ indices. \\
\phantom{x} {\bf C2:} symmetry in all lower $(\beta)$ indices.\\
\phantom{x} {\bf C3:} tracelessness  in any of it's upper $(\alpha)$ and lower $(\beta)$ indices.\\ 
Let us first consider the simpler pure irreps. of type $(n,0)$ and $(0,m)$ respectively:    
\bea 
\hspace{-2.0cm} 
|\psi\rangle ^{\alpha_1\alpha_2\ldots\alpha_n}= O^{\alpha_1\alpha_2 \ldots \alpha_n} 
|0\rangle; ~~~
|\psi\rangle_{\beta_1\beta_2\ldots\beta_m}=
O_{\beta_1\beta_2 \ldots \beta_n}|0 \rangle 
\label{symm} 
\eea
These pure representations satisfy {C1} and {C2} as the Schwinger boson creation operators 
commute amongst themselves and the condition C3 is redundant. The $(n,0)$ and $(0,m)$ Young 
tableau are given by: 
\begin{figure}[h] 
\setlength{\unitlength}{3947sp}%
\begingroup\makeatletter\ifx\SetFigFont\undefined%
\gdef\SetFigFont#1#2#3#4#5{%
  \reset@font\fontsize{#1}{#2pt}%
  \fontfamily{#3}\fontseries{#4}\fontshape{#5}%
  \selectfont}%
\fi\endgroup%
\begin{center}
\begin{picture}(3024,927)(889,-2176)
\thinlines
{\put(4401,-1561){\framebox(300,300){$\beta_m$}}
}%
{\put(3801,-1561){\framebox(600,300){$\ldots$}}
}%
{\put(901,-1561){\framebox(300,300){$\alpha_1$}}
}%
{\put(3501,-1561){\framebox(300,300){$\beta_2$}}
}%
{\put(3201,-1561){\framebox(300,300){$\beta_1$}}
}%
{\put(3201,-1861){\framebox(300,300){}}
}%
{\put(3501,-1861){\framebox(300,300){}}
}%
{\put(3801,-1861){\framebox(600,300){}}
}%
{\put(4401,-1861){\framebox(300,300){}}
}%
{\put(1201,-1561){\framebox(300,300){$\alpha_2$}}
}%
{\put(1501,-1561){\framebox(600,300){$\ldots$}}
}%
{\put(2101,-1561){\framebox(300,300){$\alpha_n$}}
}%
\put(3400,-2161){\makebox(0,0)[lb]{\smash{{\SetFigFont{12}{14.4}{\rmdefault}{\mddefault}{\updefault}{$m$ boxes $\in \,3^*$}%
}}}}
\put(1200,-1861){\makebox(0,0)[lb]{\smash{{\SetFigFont{12}{14.4}{\rmdefault}{\mddefault}{\updefault}{$n$ boxes $\in\, 3$}%
}}}}
\end{picture}%
\end{center}
\caption[N_Mirrep]{The two symmetric representations $(n,0)$ and $(0,m)$ respectively. 
 Like in SU(2) case, each single (double) box corresponds  to $a^{\dagger \alpha}$ 
($b^{\dagger}_{\alpha}$).} 
\end{figure}

\noindent Therefore, as far as symmetric representations of SU(3) are concerned, 
each single (double) 
Young tableau box represents a Schwinger boson creation operator $a^{\dagger} \in 3$ 
$\left(b^{\dagger} \in 3^*\right)$. This construction is simple and retain the 
simplicity of SU(2). However, this simplicity is lost when we consider mixed 
representations $(n \neq 0,m \neq 0)$. A $(n,m)$ representation of SU(3) has to satisfy 
C3 in addition to C1 and C2. The states in $(n,m)$ irrep. are given 
by \cite{mm1}: 
\bea
\hspace{-2.4cm} |\psi\rangle ^{\alpha_1,\alpha_2,...\alpha_n}_{\beta_1,\beta_2,...,\beta_m} \equiv
\Big[ O^{\alpha_1\alpha_2...\alpha_n}_{\beta_1\beta_2...\beta_m} + L_1 \sum_{l_1=1}^n \sum_{k_1=1}^m
\delta^{\alpha_{l_1}}_{\beta_{k_1}} O^{\alpha_1\alpha_2..\alpha_{l_1-1}\alpha_{l_1 +1}..\alpha_n}_{\beta_1\beta_2..
\beta_{k_1-1} \beta_{k_1 +1} ...\beta_m}   
 + L_2 \sum_{({}^{l_1,l_2}_{~=1})}^n \sum_{({}^{k_1,k_2}_{~=1})}^m
\delta^{\alpha_{l_1}\alpha_{l_2}}_{\beta_{k_1}\beta_{k_2}} 
\nonumber \\
\hspace{-2.4cm}O^{\alpha_1..\alpha_{l_1-1}\alpha_{l_1+1}..\alpha_{l_2-1}\alpha_{l_2+1}..\alpha_n}_{\beta_1..
\beta_{k_1-1}\beta_{k_1+1}..\beta_{k_2-1}\beta_{k_2+1} ..\beta_m}  
 + L_3 \sum_{({}^{l_1,l_2}_{l_3=1})}^n \sum_{({}^{k_1,k_2}_{k_3=1})}^m 
\delta^{\alpha_{l_1}\alpha_{l_2}\alpha_{l_3}}_{\beta_{k_1}\beta_{k_2}\beta_{k_3}} 
O^{\alpha_1..\alpha_{l_1-1}\alpha_{l_1+1}..\alpha_{l_2-1}\alpha_{l_2+1}.. \alpha_{l_3-1}\alpha_{l_3
+1}.. \alpha_n}_{\beta_1..\beta_{k_1-1}\beta_{k_1+1}..
\beta_{k_2-1}\beta_{k_2+1}..\beta_{k_3-1}\beta_{k_3+1} ...\beta_m} \nonumber \\   
\hspace{-2.3cm} +.. + L_q \sum_{l_1..l_q=1}^n \sum_{k_1..k_q=1}^{m} 
\delta^{\alpha_{l_1}\alpha_{l_2}..\alpha_{l_q}}_{\beta_{k_1}\beta_{k_2}..\beta_{k_q}}
O^{\alpha_1\alpha_2..\alpha_{l_1-1}\alpha_{l_1+1}..\alpha_{l_2-1}\alpha_{l_2+1}.. 
\alpha_{l_q-1}\alpha_{l_q+1}.. \alpha_n}_{\beta_1
\beta_2.. \beta_{k_1-1}\beta_{k_1+1}.. \beta_{k_2-1}\beta_{k_2+1}..\beta_{k_q-1}\beta_{k_q+1} ...
\beta_m} \Big] |0\rangle  
\label{bv}
\eea
where $q = {\rm min} (n,m), 
\delta^{\alpha_1\alpha_2..\alpha_r}_{\beta_1\beta_2...\beta_r} \equiv 
\delta^{\alpha_1}_{\beta_1} \delta^{\alpha_2}_{\beta_2} \ldots \delta^{\alpha_r}_{\beta_r}$ 
and all the sums in (\ref{bv}) are over different indices, i.e, 
$l_1 \neq l_2 ...\neq l_q$ and $k_1 \neq k_2 ...\neq k_q$. 
The coefficients $L_r$ are given by \cite{fn1}: 
\beq
L_r \equiv {(-1)^{r} ~(a^\dagger \cdot b^\dagger )^{r} \over {(n+m+1)
(n+m)(n+m-1)...(n+m+2-r})} ~,
\label{coef}
\eeq
The coefficients in (\ref{coef}) are chosen to satisfy the condition C3:
\beq
\sum_{i_l,j_k=1}^{3} \delta^{\alpha_l}_{\beta_k} |\psi\rangle ^{\alpha_1,\alpha_2,...\alpha_n}_{\beta_1,
\beta_2,...,\beta_m} =  0, ~~ {\rm for ~all} ~~l=1,2...n, ~~{\rm and}~~ k=1,2...m ~.
\label{trace}
\eeq
The projection operators implementing (\ref{trace}) in the Hilbert space of Schwinger bosons 
are constructed in appendix A.  
The Young tableau for the $(n,m)$ irrep. (\ref{bv}) is shown in Fig. 3. 
\begin{figure}[h]
\setlength{\unitlength}{3947sp}%
\begingroup\makeatletter\ifx\SetFigFont\undefined%
\gdef\SetFigFont#1#2#3#4#5{%
  \reset@font\fontsize{#1}{#2pt}%
  \fontfamily{#3}\fontseries{#4}\fontshape{#5}%
  \selectfont}%
\fi\endgroup%
\begin{center} 
\begin{picture}(3024,927)(889,-2176)
\thinlines
{\put(2101,-1561){\framebox(300,300){$\beta_m$}}
}%
{\put(1501,-1561){\framebox(600,300){$\ldots$}}
}%
{\put(2401,-1561){\framebox(300,300){$\alpha_1$}}
}%
{\put(1201,-1561){\framebox(300,300){$\beta_2$}}
}%
{\put(901,-1561){\framebox(300,300){$\beta_1$}}
}%
{\put(901,-1861){\framebox(300,300){}}
}%
{\put(1201,-1861){\framebox(300,300){}}
}%
{\put(1501,-1861){\framebox(600,300){}}
}%
{\put(2101,-1861){\framebox(300,300){}}
}%
{\put(2701,-1561){\framebox(300,300){$\alpha_2$}}
}%
{\put(3001,-1561){\framebox(600,300){$\ldots$}}
}%
{\put(3601,-1561){\framebox(300,300){$\alpha_n$}}
}%
\put(1276,-2161){\makebox(0,0)[lb]{\smash{{\SetFigFont{12}{14.4}{\rmdefault}{\mddefault}{\updefault}{$m$ boxes $\in \,3^*$}%
}}}}
\put(2851,-1861){\makebox(0,0)[lb]{\smash{{\SetFigFont{12}{14.4}{\rmdefault}{\mddefault}{\updefault}{$n$ boxes $\in\, 3$}%
}}}}
\end{picture}%
\end{center} 
\caption[NMirrep]{A mixed $(n,m)$ representation Young tableau. Like in SU(2) case, each 
single (double) 
box corresponds to the irreducible Schwinger boson $A^{\dagger \alpha}$ ($B^{\dagger}_{\alpha}$)
in (\ref{modsu3irrep}).} 
\end{figure} 

\noindent It is clear that  the tracelessness condition C3 makes the mixed irreps. in  
(\ref{bv}) much more involved and complicated compared to the symmetric ones (\ref{symm}). 
As a result, like in $SU(2)$ case, a chain of $n$ number of $a^\dagger$
and $m$ number of $b^\dagger$ operators acting on the vacuum does not serve the 
purpose for SU(3). In the next section, we make the construction of all SU(3) representation 
as simple as SU(2). In other words, we restore  $1-1$ correspondence between 
Young tableau boxes and the (irreducible) Schwinger boson operators. Infact, the origin of these 
problems is $a \cdot b$ and $a^{\dagger}\cdot b^{\dagger}$ which are 
SU(3) invariant operators. Another related issue is the problem of multiplicity 
\cite{mu1} arising due to the above invariants. 
Given a state $|\psi\rangle^{\alpha_1\alpha_2\ldots\alpha_n}_{\beta_1\beta_2\ldots\beta_m}$ 
in (\ref{bv}), we consider the following tower of states:  
\bea 
(a^\dagger\cdot b^\dagger)^\rho|\psi\rangle^{\alpha_1\alpha_2\ldots\alpha_n}_
{\beta_1\beta_2\ldots\beta_m} \qquad\quad  \rho=0,1,2....\infty. 
\label{jju} 
\eea
All the infinite states in (\ref{jju}) transform  like  $(n,m)$  irrep. 
as they differ by different powers of the SU(3) invariant operators. In \cite{mu1} 
it is shown that these infinite number of SU(3) degenerate states  can be uniquely labeled 
by the quantum numbers of the group Sp(2,R) which commutes with SU(3). More explicitly, if 
we define the following SU(3) invariant operators, 
$$k_{+} \equiv a^{\dagger} \cdot b^{\dagger}, ~~k_{-} \equiv a \cdot b, ~~   
	k_0 \equiv {1 \over 2} \left(N_a + N_b + 3\right),$$ 
then it is easy to check that they satisfy Sp(2,R) or SU(1,1) algebra: 
\bea 
\left[k_-,k_+\right] = 2k_0, ~~~\left[k_0,k_+\right] = k_+, ~~~\left[k_0,k_-\right] = -k_-. 
\label{sp2r} 
\eea  
It is shown in \cite{mu1} that the states in  (\ref{jju}) are in $1-1$ correspondence with the 
the positive discrete family $D_k^+$, labeled by $|k,m' \rangle$  with   
$k = {1 \over 2}(n+m+3) (={3 \over 2},2,{5 \over 2},... .)$ and  $m'= k+ \rho$. 
As $k_-|k,m'=k\rangle = 0$, the states in (\ref{bv}) are all annihilated by $k_-$, i.e: 
\bea 
k_- |\psi\rangle ^{\alpha_1,\alpha_2,...\alpha_n}_{\beta_1,\beta_2,...,\beta_m} 
= a \cdot b~ |\psi\rangle^{\alpha_1,\alpha_2,...\alpha_n}_{\beta_1,\beta_2,...,\beta_m} 
\equiv 0. 
\label{jj} 
\eea
Note that the symmetric states in (\ref{symm}) are trivially annihilated by $a \cdot b$ as they 
contain either $a^{\dagger}s$ or $b^{\dagger}s$ only. The mixed states are also annihilated 
by  $k_-$. As an example, let us consider the simplest mixed state 
$|\psi\rangle^{\alpha}_{\beta} \in 8$ representation: 
$$k_{-} |\psi\rangle^{\alpha}_{\beta} = \left(a_{\gamma}~b^{\gamma}\right) 
\left(a^{\dagger \alpha} b^{\dagger}_{\beta}-{1 \over 3} \delta^{\alpha}_{\beta} a^{\dagger} 
\cdot b^{\dagger}\right)|0\rangle = 0.$$ 
Infact,  the tracelessness condition (\ref{trace}) and the $k_-$ 
annihilation condition (\ref{jj}) are exactly equivalent (see Appendix A).  

\section{The Irreducible SU(3) Schwinger Bosons}

We have already seen that all symmetric SU(3) representations (\ref{symm}) 
retain the simplicity of SU(2).  
In this next section we define irreducible SU(3) Schwinger bosons in terms of which: 
\begin{itemize} 
\item  all mixed representations also remain as simple as SU(2) 
(compare (\ref{bv}) with  (\ref{modsu3irrep}) of this section). 
\item  the representations are multiplicity free (see eqn. (\ref{mf}) and (\ref{mf1})). 
\end{itemize} 
With the above motivation in mind, we need to define new Schwinger bosons 
$A^{\dagger \alpha}$ and $B^{\dagger}_{\alpha}$ which satisfy the following 
properties: 

\noindent (i)  $A^{\dagger}$ and $B^{\dagger}$ increase ${N}_a$ and ${N}_b$ by 1 with 
$A^{\dagger} \in 3$ and  $B^{\dagger} \in 3^*$,   

\noindent (ii) they commute amongst themselves to maintain the symmetry properties C1 and C2, 

\noindent (iii)the tracelessness property C3 is obtained by demanding the  equivalent 
(appendix A) Sp(2,R) constraint.  More explicitly: 
\bea
k_-\left(A^{\dagger\alpha} 
|\psi\rangle^{\alpha_1\alpha_2 \ldots 
\alpha_n}_{\beta_1\beta_2 \ldots \beta_m}\right)
= [k_-, A^{\dagger\alpha}] 
|\psi\rangle^{\alpha_1\alpha_2 \ldots 
\alpha_n}_{\beta_1\beta_2 \ldots \beta_m} =0. 
\label{AA}
\eea
The most general form of $A^\dagger_\alpha$ 
consistent with (i) and (iii) is: 
\bea
\label{form_A}
A^{\dagger\alpha}= 
a^{\dagger\alpha}+ {\mathrm L}({N}_a,{N}_b) k_+b^\alpha
\eea
Now,  ${\mathrm L}({N}_a,{N}_b)$ is fixed by:
\bea
 k_- A^{\dagger\alpha} 
|\psi\rangle^{\alpha_1 \alpha_2 \ldots \alpha_n}_{\beta_1 \beta_2 
\ldots \beta_m} = \left(1 + 
(n+m+3) {\mathrm L}(n+1,m+1)\right) b^\alpha|\psi\rangle^{\alpha_1 \alpha_2 
\ldots \alpha_n}_{\beta_1 \beta_2 \ldots \beta_m} = 0 
\eea
In (\ref{abcd}) we have made use of (\ref{sp2r}) and (\ref{jj}). 
This fixes 
\bea 
{\mathrm L}(n,m)= -{1 \over (n+m+1)}
\eea 
and we get: 
\bea
A^{\dagger\alpha}= a^{\dagger\alpha}-\frac{1}{{N}_a+{N}_b+1}k_+b^\alpha, ~~~~
A_\alpha&=& a_\alpha-b^\dagger_\alpha k_-\frac{1}{{N}_a+{N}_b+1}  
\label{abcd} 
\eea
Similarly, 
\bea
B^\dagger_\alpha= b^\dagger_\alpha-\frac{1}{{N}_a+{N}_b+1}k_+a_\alpha, ~~~~~~
B^\alpha &=& b^\alpha-a^{\dagger{\alpha}} k_-\frac{1}{{N}_a+{N}_b+1}
\eea
It is easy to check that the irreducible Schwinger boson creation operators 
commute amongst themselves: 
\bea 
\left[A^{\dagger \alpha}, A^{\dagger \beta}\right] = 0, ~~ \left[B^{\dagger}_{\alpha}, 
B^{\dagger}_{\beta}\right] = 0, ~~ \left[A^{\dagger \alpha}, B^{\dagger}_{\beta}\right] =0.   
\label{comm} 
\eea
The other commutation relations acting on the SU(3) irreps. are: 
\bea 
\left[A_{\alpha}, A^{\dagger \beta}\right]
|\psi\rangle^{\alpha_1\alpha_2 \ldots 
\alpha_n}_{\beta_1\beta_2 \ldots \beta_m}
& = &  
\left(\delta_{\alpha}^{\beta} - {1 \over {N_a+N_b+2}} B^{\dagger}_{\alpha}B^{\beta}\right) 
|\psi\rangle^{\alpha_1\alpha_2 \ldots 
\alpha_n}_{\beta_1\beta_2 \ldots \beta_m}, \nonumber \\ 
\label{commr} 
\left[A_{\alpha}, B^{\dagger}_{\beta}\right]  
|\psi\rangle^{\alpha_1\alpha_2 \ldots 
\alpha_n}_{\beta_1\beta_2 \ldots \beta_m}
& = & ~~-{1 \over {N_a+N_b+2}} B^{\dagger}_{\alpha} A_{\beta} 
|\psi\rangle^{\alpha_1\alpha_2 \ldots 
\alpha_n}_{\beta_1\beta_2 \ldots \beta_m},
\\ 
\left[B^{\alpha}, B^{\dagger}_{\beta}\right] 
|\psi\rangle^{\alpha_1\alpha_2 \ldots 
\alpha_n}_{\beta_1\beta_2 \ldots \beta_m}
& = & 
\left(\delta^{\alpha}_{\beta} - {1 \over {N_a+N_b+2}} A^{\dagger \alpha}A_{\beta}\right) 
|\psi\rangle^{\alpha_1\alpha_2 \ldots 
\alpha_n}_{\beta_1\beta_2 \ldots \beta_m}.  \nonumber   
\eea

\subsection{\bf SU(3) Representations} 

Hence a general $(n,m)$ irreducible representation of $SU(3)$ can be written in 
terms of these irreducible Schwinger bosons as:
\bea
\label{modsu3irrep}
\hspace{-1.6cm} |{{\Psi}}\rangle^{\alpha_1\alpha_2\ldots\alpha_n}_{\beta_1\beta_2\ldots\beta_m}
\equiv {\bf{O}}^{\alpha_1\alpha_2\ldots  \alpha_n}_{\beta_1\beta_2 \ldots \beta_m}
|0\rangle
= A^{\dagger\alpha_1}A^{\dagger\alpha_2}\ldots A^{\dagger\alpha_n}
B^{\dagger}_{\beta_1}B^{\dagger}_{\beta_2}\ldots B^{\dagger}_{\beta_m}
|0\rangle
\eea 
The simple construction (\ref{modsu3irrep}) is equivalent to (\ref{bv}). To see the equivalence, 
it is instructive to first consider some simple examples. The two simplest fundamental 
representations ($(1,0)$ and $(0,1)$) are: 
\bea
\label{(0,1)}
|\Psi\rangle^\alpha
=A^{\dagger\alpha}|0\rangle=\left( a^{\dagger\alpha}-
\frac{1}{{N}_a+{N}_b+1}k_+b^\alpha \right)|0\rangle=a^{\dagger\alpha}|0\rangle \equiv 
|\psi\rangle^\alpha \nonumber \\
|\Psi\rangle_{\beta}= ~B^\dagger_\beta|0\rangle=\left( b^\dagger_\beta-
\frac{1}{{N}_a+{N}_b+1}k_+a_\beta \right)|0\rangle=b^\dagger_\beta|0\rangle  
\equiv |\psi\rangle_\beta.
\eea
The equations (\ref{(0,1)}) also demonstrate that all the  
symmetric (i.e $(n,0),(0,m)$) representations in terms of irreducible Schwinger 
boson are exactly same as before. 
This is of course trivial. The simplest mixed $(1,1)$ states are: 
\bea 
\hspace{-1.2cm} |\Psi\rangle^{\alpha}_{\beta} & \equiv &  A^{\dagger \alpha} B^{\dagger}_{\beta} 
| 0 \rangle  =  
\left(a^{\dagger \alpha}-\frac{1}{{N}_a+{N}_b+1}k_+b^\alpha\right) 
\left(b^\dagger_\beta-\frac{1}{{N}_a+{N}_b+1}k_+a_\beta\right) |0\rangle  \nonumber \\
&=& \left(a^{\dagger \alpha} b^{\dagger}_\beta -
\frac{\delta^{\alpha}_{\beta}}{3} (a^{\dagger}\cdot b^{\dagger})\right) |0\rangle \equiv 
|\psi\rangle^{\alpha}_{\beta}.
\label{y1y2} 
\eea 
\noindent The equivalence between (\ref{bv}) and (\ref{modsu3irrep}) is  
explicitly established using the method of induction in Appendix B. 
Infact, it is easy to see that the states (\ref{modsu3irrep}) satisfy 
all the three conditions: C1, C2 and C3 mentioned in section III.
The symmetry properties C1 and  C2 of (\ref{modsu3irrep}) follow 
from the commutation relations (\ref{comm}). The tracelessness property C3 of all the  
mixed states $(n,m)$ with $n,m=1,2\ldots \infty$  in (\ref{modsu3irrep}) also follows 
(infact obvious) from the tracelessness of the octet state (\ref{y1y2}). To see this, we consider: 
\bea 
|\Psi\rangle^{\gamma\alpha_2 \ldots \alpha_n}_{\gamma\beta_2\ldots \beta_m} =  A^{\dagger} \cdot B^{\dagger}   
|\Psi\rangle^{\alpha_2 \ldots \alpha_n}_{\beta_2 \ldots \beta_m}  
= A^{\dagger\alpha_2}\ldots A^{\dagger\alpha_n}
B^{\dagger}_{\beta_2}\ldots B^{\dagger}_{\beta_m}
|\Psi\rangle^{\gamma}_{\gamma}  =0 
\label{mf}
\eea 
In (\ref{mf}), we have used the fact that all the $A^{\dagger}s$ 
and $B^{\dagger}s$ commute amongst themselves (\ref{comm}) and the octet state (\ref{y1y2}) 
is traceless. 
Therefore, the SU(3) $(n,m)$ representation states in (\ref{modsu3irrep}) are exactly 
same as the states in (\ref{bv}):
$|{{\Psi}}\rangle^{\alpha_1\alpha_2\ldots\alpha_n}_{\beta_1\beta_2\ldots\beta_m}
\equiv |{{\psi}}\rangle^{\alpha_1\alpha_2\ldots\alpha_n}_{\beta_1\beta_2\ldots\beta_m}.$ 
We further note that: 
\bea
A \cdot B~  
|\Psi\rangle^{\alpha_1\alpha_2 \ldots 
\alpha_n}_{\beta_1\beta_2 \ldots \beta_m}  = 
a \cdot b~ 
|\Psi\rangle^{\alpha_1\alpha_2 \ldots 
\alpha_n}_{\beta_1\beta_2 \ldots \beta_m} = 0. 
\label{mf1} 
\eea
Hence,  the present construction in terms of irreducible Schwinger bosons also solves the 
multiplicity problem. The eqns. (\ref{mf}) and (\ref{mf1}) show that it is no longer possible 
to construct the infinite tower (\ref{jju}) in terms of irreducible Schwinger bosons. 
We again emphasize that each Young tableau single (double) box $\in 3$ 
($3^*$) representation in Figure 3  corresponds to the irreducible Schwinger boson creation 
operator $A^{\dagger \alpha}$ ($B^{\dagger \alpha}$). This SU(3) representation 
feature is again like SU(2) representations.  

\section{Summary \& Discussion} 

\noindent We conclude that the irreducible Schwinger bosons make the SU(3) 
representation theory as simple as SU(2). By constructing irreducible 
Schwinger bosons commuting with $k_-$, we are able to produce all the SU(3) irreps. 
with the ease of SU(2). Infact, the irreducible Schwinger bosons play 
important physical role in lattice gauge theories. In SU(2) lattice gauge theories 
the SU(2) Schwinger bosons create the electric or angular momentum fluxes along 
the links \cite{lgt}. In SU(3) lattice gauge theory, the corresponding 
role is played by irreducible SU(3) Schwinger bosons \cite{rmi}. 

We now briefly discuss the extension of the ideas in this paper to the 
SU(N)  group. The SU(N) generalization of the SU(2) Schwinger 
mapping was done in \cite{mm2} in the context of SU(N) 
coherent states. The SU(N) group has rank $(N-1)$. 
Therefore, we now define the SU(N) generators as: 
\bea
Q^{a} = \sum_{r=1}^{N-1} Q^{a}[r] = \sum_{r=1}^{N-1} 
a^{\dagger}[r] \frac{\lambda^{a}[r]}{2} a[r].
\label{sbnn}
\eea
In (\ref{sbnn}), the index  $r(=1,2,...,(N-1))$ covers all the $(N-1)$ fundamental 
representations of SU(N) group. The invariance group is now bigger than Sp(2,R) and 
can again be used to define irreducible SU(N) Schwinger bosons. This will again lead 
to a simplified representation theory of SU(N). The work in this direction 
is in progress and will be reported elsewhere. \\ 

\noindent {\it Acknowledgment:} The authors would like to thank H. S. Sharatchandra for 
useful discussions. 

\appendix
\section{SU(3) Projection Operators} 
We  now construct projection operators $P(n,m)$ such that,
\bea
\label{prj}
P_{(n,m)}
~O^{\alpha_1\alpha_2 \ldots \alpha_n}_{\beta_1\beta_2 \ldots \beta_m}
|0 \rangle = 
|\psi\rangle^{\alpha_1\alpha_2 \ldots \alpha_n}_{\beta_1\beta_2 \ldots \beta_m} 
\eea
where, $O^{\alpha_1\alpha_2 \ldots \alpha_n}_{\beta_1\beta_2 \ldots \beta_m}$ 
is defined in (\ref{defo}). 
It is clear that the state 
$|\psi\rangle^{\alpha_1\alpha_2 \ldots \alpha_n}_{\beta_1\beta_2 \ldots \beta_m}$ 
will transform in the same way as 
$O^{\alpha_1\alpha_2 \ldots \alpha_n}_{\beta_1\beta_2 \ldots \beta_m} 
|0\rangle$. Hence the projection operator can contain only $SU(3)$ invariant 
operators. Thus the most general form of $P_{(n,m)}$ is given by: 
\bea 
P_{(n,m)} \equiv \sum_{r=0}^{\infty} l_r(n,m) (k_+)^r (k_-)^r =\sum_{r=0}^{q} l_r(n,m) (k_+)^r (k_-)^r 
\eea 
where, $q=\mbox{min}(n,m)$.\\
Applying $k_-$ on (\ref{prj}) and equating it to zero, we get the recurrence 
relation: 
\bea 
\frac{(n+ m+2-r)}{(r-1)}l_r(n,m)=-l_{r-1}(n,m)
\label{xyz} 
\eea 
Choosing the overall normalization $l_{0}=1$, the solution of (\ref{xyz}) 
is 
\bea 
\hspace{-1.6cm} l_r(n,m)=\frac{(-1)^r}{r!(n+m+1)\ldots(n+m+2-r)}= 
\frac{(-1)^r}{r!}\frac{( n+m+1-r)!}{( n+ m+1)!}, 
\label{xbv} 
\eea
leading to: 
\bea
\label{P0} 
P_{(n,m)} = \frac{1}{(n+m+1)!} \sum_{r=0}^{\infty} 
\frac{(-1)^r}{r!}({n}+{m}+1-r)! (k_+)^r (k_-)^r 
\eea 
The action of the projection operator on the state $O^{\alpha_1\alpha_2 \ldots 
\alpha_n}_{\beta_1\beta_2 \ldots \beta_m} |0 \rangle$ leads to (\ref{bv}). Infact 
the projection operator in (\ref{P0}) is idempotent, i.e, it 
satisfies: 
\bea
\label{idem}
P_{(n,m)}|\psi\rangle^{\alpha_1 \alpha_2 
\ldots \alpha_n}_{\beta_1 \beta_2 \ldots \beta_m }
=|\psi\rangle^{\alpha_1 \alpha_2 \ldots 
\alpha_n}_{\beta_1 \beta_2 \ldots \beta_m}
\eea
The above property is obvious as $k_-$ annihilates the states
$|\psi\rangle^{\alpha_1 \alpha_2 \ldots 
\alpha_n}_{\beta_1 \beta_2 \ldots \beta_m}$ and therefore only 
the identity (i.e,  $r=0$ term) in (\ref{P0}) contributes. 
 
\section{Action of Irreducible Schwinger bosons}

In this appendix we  show that the states in (\ref{modsu3irrep}) 
are same as the SU(3) irreps. in (\ref{bv}). These 
results are obvious for all the symmetric representations 
as shown in section IV A We have also seen this equivalence for 
the octet $(1,1)$ representation. We now use the method of 
induction for the general case.  Let us assume the equivalence 
for $(n,m)$ representation: 
\bea  
|{{\Psi}}\rangle^{\alpha_1\alpha_2\ldots\alpha_n}_{\beta_1\beta_2\ldots\beta_m}
\equiv 
|{{\psi}}\rangle^{\alpha_1\alpha_2\ldots\alpha_n}_{\beta_1\beta_2\ldots\beta_m}.
\label{xx1} 
\eea 
We now need to prove:
\bea
\label{st}
|\Psi\rangle^{\alpha_1\alpha_2\ldots\alpha_{n+1}}_{  
\beta_1 \beta_2 \ldots \beta_{m+1}}
= |\psi\rangle^{\alpha_{}\alpha_1\alpha_2\ldots\alpha_{n+1}}_{\beta_{} 
\beta_1 \beta_2 \ldots \beta_{m+1}} 
\eea
Let us first consider the case $n \rightarrow n+1$ and $m \rightarrow m$. The l.h.s.
of (\ref{st}) is: 
\bea
 |\Psi\rangle^{\alpha\alpha_1\alpha_2\ldots\alpha_n}_{\beta_1 \beta_2 \ldots \beta_m}
  & \equiv &      
A^{\dagger{\alpha}} |\Psi\rangle^{\alpha_1 \alpha_2 \ldots \alpha_n}_{\beta_1 \beta_2 \ldots \beta_m}
 =   A^{\dagger{\alpha}} |\psi\rangle^{\alpha_1 \alpha_2 
\ldots \alpha_n}_{\beta_1 \beta_2 \ldots \beta_m}\nonumber \\
& =& \left(a^{\dagger{\alpha}}-\frac{1}{N_a+N_b+1}k_+b^{\alpha}\right) \left(1 + 
\sum_{r=1}^{\infty} l_r(n,m)k_+^rk_-^r\right)O^{\alpha_1 \alpha_2 \ldots \alpha_n}_{\beta_1 \beta_2 
\ldots \beta_m}|0\rangle \nonumber \\
& =& \Big(\underbrace{O^{\alpha\alpha_1\alpha_2\ldots\alpha_n}_{\beta_1 \beta_2 
\ldots \beta_m}}_{T_1}  - \underbrace{\frac{k_+b^\alpha}{(n+m+2)}
O^{\alpha_1\alpha_2\ldots\alpha_n}_{\beta_1 \beta_2 \ldots \beta_m}}_{T_2}  
+\underbrace{\sum_{r=0}^{\infty} l_r(n,m)
k_+^r a^{\dagger\alpha}k_-^r O^{\alpha_1\alpha_2\ldots\alpha_n}_{\beta_1 \beta_2 \ldots \beta_m}}_{T_3} 
\nonumber \\&{}&  \hspace{0.8cm} -\underbrace{\frac{1}
{(n+m+2)} \sum_{r=0}^{\infty} l_r(n,m)k_+b^\alpha k_+^rk_-^r 
O^{\alpha_1\alpha_2\ldots\alpha_n}_ {\beta_1 \beta_2 \ldots \beta_m}
}_{T_4}\Big) \,|0\rangle  
\label{terms}
\eea
We use:
$$[a^{\dagger\alpha},k_-^r]=-rk_-^{r-1}b_\alpha, ~~~ [b^{\alpha},k_+^r]=rk_+^{r-1}a^{\dagger \alpha},$$ 
to write the third term ($T_3$) and the fourth term ($-T_4$) in (\ref{terms}) as: 
\bea
T_3  & = &  \underbrace{\sum_{r=1}^{\infty} l_r(n,m) k_-^rk_+^r 
O^{\alpha \alpha_1\alpha_2\ldots\alpha_n}_ {\beta_1 \beta_2 \ldots \beta_m}}_{T_{31}}  
- \underbrace{\sum_{r=1}^{\infty} 
r l_r(n,m) k_+^r k_-^{r-1} b^{\alpha}   
O^{\alpha_1\alpha_2\ldots\alpha_n}_ {\beta_1 \beta_2 \ldots \beta_m}}_{T_{32}} \nonumber \\
T_4  & = &   
\underbrace{\sum_{r=1}^{\infty} \frac{r l_r(n,m)}{n+m+2} k_+^rk_-^ra^{\dagger \alpha} 
O^{\alpha_1\alpha_2\ldots\alpha_n}_ {\beta_1 \beta_2 \ldots \beta_m}}_{T_{41}} 
- \underbrace{\sum_{r=1}^{\infty} \frac{r^2 l_r(n,m)}{n+m+2} k_+^rk_-^{r-1} b^{\alpha} 
O^{\alpha_1\alpha_2\ldots\alpha_n}_ {\beta_1 \beta_2 \ldots \beta_m}}_{T_{42}} 
\nonumber \\
& & +  \underbrace{\sum_{r=1}^{\infty} \frac{l_r(n,m)}{n+m+2} k_+^{r+1} k_-^{r} b^{\alpha} 
O^{\alpha \alpha_1\alpha_2\ldots\alpha_n}_ {\beta_1 \beta_2 \ldots \beta_m}}_{T_{43}}  
\eea 
The defining eqn. (\ref{xbv}) implies: 
\bea 
l_r(n+1,m) \equiv \frac{n+m+2-r}{n+m+2}l_r(n,m). 
\label{cnv} 
\eea
Using (\ref{cnv}), (\ref{P0}) and (\ref{prj}), we get: 
\bea 
\left(T_1+T_{31}-T_{41}\right) |0\rangle   =  
|\psi\rangle^{\alpha\alpha_1 \ldots \alpha_n}_{\beta_1\beta_2 \ldots \beta_m}. 
\label{finn} 
\eea   
We now need to show: 
$$-T_2+T_{32}-T_{42}-T_{43} = 0.$$  
Using (\ref{cnv}) and  
\bea 
- \frac{l_r(n,m)}{n+m+2} =  (r+1) l_{r+1}(n+1,m), ~~~ l_1(n+1,m) = \frac{-1}{n+m+2}, \nonumber 
\eea 
we get: 
\bea 
T_{32}-T_{42}  & = &  
- \sum_{r=1}^{\infty} r l_r(n+1,m) k_+^r k_-^{r-1} b^{\alpha} 
O^{\alpha_1\alpha_2\ldots\alpha_n}_{\beta_1 \beta_2 \ldots \beta_m} \\   
-T_2-T_{43}  & = &  \left(\sum_{r=1}^{\infty} (r+1) l_{r+1}(n+1,m) k_+^{r+1}k_-^r  
+ l_1(n+1,m) k_+\right) b^\alpha  
O^{\alpha_1\alpha_2\ldots\alpha_n}_{\beta_1\beta_2 \ldots \beta_m} \nonumber \\
& = & \sum_{r=1}^{\infty} r l_r(n+1,m) k_+^r k_-^{r-1} b^{\alpha} 
O^{\alpha_1\alpha_2\ldots\alpha_n}_{\beta_1 \beta_2 \ldots \beta_m}  \equiv T_{42}-T_{32}   
\label{nmj} 
\eea
Similarly, we can prove the case: $m \rightarrow m+1$ and $n \rightarrow n$. Thus we have also 
explicitly proved that the simple SU(3) Schwinger boson states (\ref{modsu3irrep}) (which are 
exact SU(3) analogues of the SU(2) construction (\ref{su2_irrep})) are indeeed
the SU(3) irrep. states (\ref{bv}).



\begin{thebibliography}{99}
\bibitem{fn1} The expression for $L_r$ in \cite{mm1} is wrong, it contains an extra 
factor ${1 \over r!}$.
\end{thebibliography}
\end{document}